\documentclass[runningheads]{llncs}

\usepackage{wrapfig}

\usepackage{soul,times}

\soulregister\overline7
\soulregister\cite7
\soulregister\ref7
\soulregister\dyn7

\usepackage{balance}
\usepackage{todonotes}
\usepackage{amsmath,amssymb,amsfonts}
\usepackage{algorithmic}
\usepackage{algorithm}
\usepackage{graphicx}
\usepackage{textcomp}
\usepackage{url,multirow,fontawesome}
\usepackage{xcolor}
\usepackage{tcolorbox}
\usepackage{enumitem}

\colorlet{darkpink}{red!60!purple!50}
\colorlet{lightpurple}{purple!60!blue!60}
\colorlet{purple}{purple!70!blue!100}
\colorlet{teal}{green!70!blue!90!black!100}
\colorlet{darkgreen}{green!50!black!100}
\colorlet{orange}{orange!70!red!80}
\colorlet{yellow}{yellow!100!orange!70!red!80!black!80}

\newcommand{\TT}[1]{\texttt{#1}}

\usepackage{stmaryrd}
\usepackage{color}
\usepackage{float}
\usepackage{xfrac}
\usepackage{bbold}
\usepackage{url}
\usepackage{graphicx}
\usepackage{caption}
\usepackage{subcaption}
\usepackage{xspace}
\usepackage{mathtools}
\usepackage{mathpartir}
\usepackage{bm}
\usepackage{xcolor}
\usepackage{comment}

\usepackage{amsthm}

\theoremstyle{plain}

\theoremstyle{definition}

\usepackage{listings}

\usepackage[T1]{fontenc}
\usepackage{microtype}

\newcommand{\x}{x}

\newcommand{\match}{\textbf{match}}
\newcommand{\tlet}{\textbf{let}}
\newcommand{\rec}{\textbf{rec}}
\newcommand{\tin}{\textbf{in}}
\newcommand{\with}{\textbf{with}}
\newcommand{\dyn}{\texttt{?}}

\begin{document}
\title{Putting gradual types to work}
%
%
\author{ Bhargav Shivkumar\inst{1,2} \orcidID{0000-0002-8430-9229} \and
Enrique Naudon\inst{1}\orcidID{0000-0001-9878-2781} \and Lukasz Ziarek\inst{2} }
\authorrunning{Shivkumar et al.}
\institute{Bloomberg, New York, USA \and SUNY - University at Buffalo , New York, USA}
\maketitle              
\begin{abstract}
In this paper, we describe our experience incorporating gradual types in a statically typed functional language with Hindley-Milner style type inference. Where most gradually typed systems aim to improve static checking in a dynamically typed language, we approach it from the opposite perspective and promote dynamic checking in a statically typed language. Our approach provides a glimpse into how languages like SML and OCaml might handle gradual typing.  We discuss our implementation and challenges faced---specifically how gradual typing rules apply to our representation of composite and recursive types. We review the various implementations that add dynamic typing to a statically typed language in order to highlight the different ways of mixing static and dynamic typing and examine possible inspirations  while maintaining the gradual nature of our type system. This paper also discusses our motivation for adding gradual types to our language, and the practical benefits of doing so in our industrial setting.

\end{abstract}

\def\CC{{C\nolinebreak[4]\hspace{-.05em}\raisebox{.2ex}{\small\bf ++}}}

\section{Introduction}
\label{sec:intro}
Static typing and dynamic typing are two opposing type system paradigms. Statically typed languages are able to catch more programmer bugs early in the compilation process, at the expense of a more flexible semantics.  On the other hand, dynamically typed languages allow greater flexibility, while allowing more bugs at runtime. The proponents of each paradigm often feel very strongly in favor of their paradigm.  Language designers are stranded in the middle of this dichotomy and left to decide between the two extremes when designing their languages.

At Bloomberg, we have felt this pain while designing a domain specific language  for programmatically defining financial contracts.  For the purposes of this paper, we will call our language Bloomberg Contract Language (BCL).  BCL is a statically typed functional language with Hindley-Milner style type inference~\cite{damas1982principal,milner1978theory}, structural composite types and recursive types.  Users of BCL are split into two groups---end users and language maintainers. End users are typically financial professionals whose primary programming experience involves scripting in dynamically typed languages such as Python and MATLAB. On the other hand, language maintainers are Bloomberg software engineers who are most at ease programming in statically typed and often functional languages like OCaml. Whilst it is of paramount importance to provide our end users with an environment in which they are comfortable, our domain---financial contracts---is one in which correctness is of extraordinary importance, since errors can lead to large financial losses.  This makes static types appealing, as they catch many errors that dynamic systems might miss. Even though static types provide a more error-free runtime, they do require extra effort from our end users who must learn an unfamiliar system.  Our desire to simultaneously satisfy our end users and our language maintainers led us to gradual typing~\cite{siek2006gradual}, which seeks to integrate static and dynamic typing in one system.  Gradual typing in BCL allows language maintainers to stick to static typing and end users to selectively disable static typing when it interferes with their ability to work in BCL.

Since its introduction, gradual typing~\cite{siek2006gradual} has been making its way into more mainstream languages~\cite{10.1145/2661088.2661101,tobin2008design} and more people have acknowledged the varied benefits of mixing static and dynamic typing in the same program. As identified by Siek and Taha~\cite{siek2015criteria}, there has been considerable interest in integrating static and dynamic typing, both in academia and in industry.  There has also been a plethora of proposed approaches, from adding a dynamic keyword~\cite{abadi1991dynamic}, to using objects in object-oriented languages~\cite{meijer2004static}, to Seik and Taha's gradual typing itself~\cite{siek2006gradual}. While there seems to be no one-size-fits-all approach to designing a system that mixes static and dynamic types, Siek and Taha standardize the guarantees~\cite{siek2015criteria} we can expect from such a system. For language designers, this provides a more methodical way to approach the integration. Language designers can also draw from a large body of literature exploring the combination of gradual types with other common features, such as objects~\cite{siek2007gradual} and type inference~\cite{siek2008gradual,garcia-cimini-2015}.

While it is typical for dynamically typed languages to go the gradual route in order to incorporate more static type checking, we go the other way and add more dynamism to our already static language. Most static languages that incorporate dynamic typing do so by following in the footsteps of Abadi et.al.~\cite{abadi1991dynamic}--C\# is a prime example of this~\cite{hejlsberg2010c}. Since BCL already supports type inference and we want to retain the dynamic feel of the language, we implement the inference algorithm described by Siek and Vachhrajani~\cite{siek2008gradual}, putting us in an interesting position. Our approach promotes the use of a $\dyn$ annotation to explicitly signify dynamically typed terms while un-annotated terms are (implicitly) statically typed, much like that of Garcia and Cimini~\cite{garcia-cimini-2015}. This approach provides a simple escape hatch to end users who want to use dynamic typing as well as avenues to automate this process to ensure backwards compatibility of BCL with legacy code. 

Finally, we feel there is a need to study the adaptation of gradual types to an existing language with a substantial user base and lots of existing code. We aim to provide a technical report in this paper that models our design decisions and implementation details of bringing in gradual types to BCL. Our primary contributions include: 
\begin{itemize}
    \item A brief review of other statically typed languages that add dynamic types, to compare and possibly derive inspiration for our own design in Section~\ref{sec:background}.
    \item Introduce a new use case that shows how a gradually typed language benefits different user groups of a language in Section~\ref{sec:bcl}.
    \item An inference algorithm, which is an adaptation of a prominent inference algorithm to add gradual types to a language with type inference in Section~\ref{sec:implementation}.
\end{itemize}
Note that throughout this paper we use "gradual" to indicate an implementation that provides gradual guarantees as specified in~\cite{siek2015criteria}. While, we do not state this formally for BCL and leave that to future work, our implementation supports the smooth evolution of programs from static to dynamic typing as prescribed for gradually typed systems.

\section{Background}
\label{sec:background}
In this section we briefly survey the existing literature to better contextualize our design choices.
The incorporation of static and dynamic typing has been extensively studied~\cite{tobin2006interlanguage,matthews2009operational,siek2006gradual,siek2015criteria,gronski2006sage}, though usually in the context of a core calculus instead of a full-featured language. There also seems to be a juxtaposition of the literature, which generally follows a static-first approach, and practical implementations, which generally follow a dynamic-first approach~\footnote{Here, static-first refers elaborating a static surface language to a gradually typed intermediate representation.  Conversely, by dynamic-first we mean the opposite: elaborating a dynamic surface language to a gradually typed intermediate representation.}~\cite{greenberg2019dynamic}.  

Abadi et al~\cite{abadi1991dynamic} has been an inspiration for many static languages looking to incorporate dynamic typing. This work is a precursor to gradual typing, and while it does not qualify as gradual à la~\cite{siek2006gradual}, it is nevertheless a standard when it comes to adding dynamic checks to a static language. Abadi's work uses a \TT{dynamic} construct to build terms of type \TT{Dynamic} and a \TT{typecase} construct to perform case analysis on the runtime type of an expression of type \TT{Dynamic}. This is similar to the \TT{typeof()} function in dynamic languages like Python, which resolve the type of an expression at runtime. Siek and Taha observe that translating from their language of explicit casts to Abadi et al's language is not straightforward~\cite{siek2006gradual}.  Nevertheless we believe that it is worthwhile to introduce something like the \TT{typecase} construct in a static language with gradual types.  We identify and discuss some potential applications of this in Section~\ref{sec:benefits}.

Statically typed object oriented languages like C\# and Java have worked to incorporate some form of dynamic typing~\cite{gray2005fine,meijer2004static}. C\# 4.0 introduced the \TT{dynamic} type to declare objects that can bypass static type checking~\cite{csharpdynamic}. Although this achieves dynamic type checking, there is no indication of it being gradual à la~\cite{siek2006gradual}. Moreover, using the \TT{dynamic} type in a C\# program runs the program on the Dynamic Language Runtime (DLR) which is a separate runtime from the Common Language Runtime and which supports dynamic checking.

While works like~\cite{10.1145/3290331,10.1145/2103621.2103714} examine gradual type inference from the perspective of removing dynamic checks by performing type inference at runtime, Garcia and Cimini~\cite{garcia-cimini-2015} (much like BCL) deals with static reasoning about programs, based on the consistency relation.~\cite{garcia-cimini-2015} explores an alternate approach to gradual type inference and presents a statically typed language and its gradual counterpart. Instead of inferring gradual types based on type precision~\cite{siek2008gradual}, this work limits the inference problem to static types only and requires consistency constraints between gradual types. An interesting feature of their language is that they distinguish between static type parameters and gradual type parameters to tell static parametric polymorphism apart from polymorphism due to the dynamic type.

Our approach is to adopt the properly gradual system defined by Siek and Vachchrajani~\cite{siek2008gradual}. That work describes the incorporation of gradual typing into a language with unification-based type inference.  Unification-based inference is a common implementation of the Hindley-Milner type system~\cite{milner1978theory}, and is the implementation that BCL already uses. This makes our integration work relatively easier and also lets us leverage all the benefits of the standard for gradual typing laid out by Siek and Taha~\cite{siek2015criteria}.

\subsection{Gradual types and unification based inference}

\begin{figure}[ht]
   \centering
	\subfloat[$\lambda_\rightarrow^\alpha$\label{fig:lam-a-ts}]{
\resizebox{0.48\columnwidth}{!}{%
      \begin{tabular}[width=.5\textwidth]{lcl}
            \quad (\textsc {SVAR}) & \inferrule{\Gamma (\x) = \tau}{S; \Gamma \vdash \x : \tau} & \boxed{S;\Gamma \vdash e : \tau } \\\\
            \quad (\textsc {SCNST}) & \inferrule{S;\Gamma \vdash c : \emph{typeof}(c)}{} &  \\\\
            \quad (\textsc {SAPP}) & \inferrule{S;\Gamma \vdash e_1 : \tau_1 \\
                                               S;\Gamma \vdash e_2 : \tau_2 \\\\
                                                S(\tau_1) = S(\tau_2 \rightarrow \tau_3)}
                                             {S;\Gamma \vdash e_1 \ e_2 : \tau_3}  & \\\\
             \quad (\textsc {SABS}) & \inferrule
            {S;\Gamma(\x \mapsto \tau_1) \vdash e : \tau_2}
            {S;\Gamma \vdash \lambda\x : \tau_1 . e : \tau_1 \rightarrow \tau_2} &\\\\
      \end{tabular}%
      }
	}
	\subfloat[$\lambda_\rightarrow^{?\alpha}$ \label{fig:lam-dyn-a-ts}]{
\resizebox{0.48\columnwidth}{!}{%
	\begin{tabular}[width=.5\textwidth]{lcl}
        \quad (\textsc {GVAR}) & \inferrule{\Gamma (\x) = \tau}{S;\Gamma \vdash_g \x : \tau} & \boxed{S;\Gamma \vdash_g e : \tau} \\\\
        \quad (\textsc {GCNST}) & \inferrule{S;\Gamma \vdash_g c : \emph{typeof}(c)}{} & \\\\
        \quad (\textsc {GAPP}) & \inferrule{S;\Gamma \vdash_g e_1 : \tau_1 \\ S; \Gamma \vdash_g e_2 : \tau_2  \\\\ S \models \tau_1 \simeq \tau_2 \rightarrow \beta \\ (\beta fresh)}{S;\Gamma \vdash_g e_1 \ e_2 : \beta}
         & \\\\
        \quad (\textsc {GABS}) & \inferrule{S;\Gamma(\x \mapsto \tau_1) \vdash_g e : \tau_2}{S;\Gamma \vdash_g \lambda\x :\tau_1 . e : \tau_1 \rightarrow \tau_2}
        & \\\\
    \end{tabular}%
    }
}
	\caption{Simply and gradually typed lambda calculus with type variables}
\end{figure}

\begin{figure}[h]
	\centering
    \includegraphics[width = \textwidth]{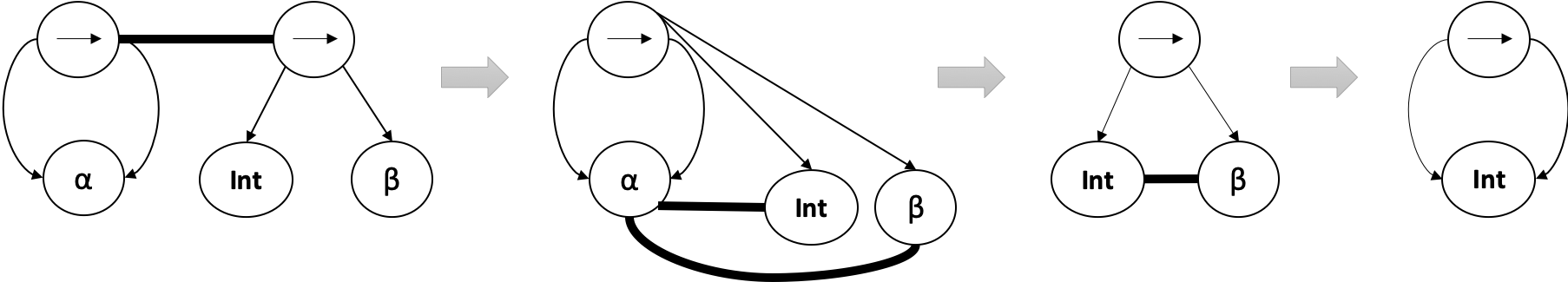}
	\caption{Huet's unification of \( \{ \alpha \rightarrow \alpha = Int \rightarrow \beta  \} \)}
	\label{fig:huet}
\end{figure}

Siek and Vachchrajani~\cite{siek2008gradual}(S\&V) propose an innovative solution for performing gradual type inference which combines gradual typing with type inference. Their main goal is to allow inference to operate on the statically typed parts of the code, while leaving the dynamic parts to runtime checks.  Furthermore, the dynamic type must unify with static types and type variables, so that the static and dynamic portions of code may freely interact.  In this section, we summarize their work.

The work of S\&V is based on the gradually typed lambda calculus~\cite{siek2006gradual}.  The gradually typed lambda calculus extends the simply typed lambda calculus (\(\lambda_\rightarrow\)) with an unknown type, \(?\)--pronounced ``dynamic''; type checking for terms of this type is left until runtime.  The gradually typed lambda calculus  (\(\lambda_\rightarrow^?\)) allows static and dynamic types to freely mix and satisfies the gradual guarantee~\cite{siek2015criteria}, ensuring smooth migration between static and dynamic code while maintaining the correctness of the program.

Type inconsistencies in \(\lambda_\rightarrow^?\) are caught by a \emph{consistent} relation, instead of equality as in \(\lambda_\rightarrow\).  The \emph{consistent} relation only compares parts of a type that are statically known; it is one of the key contributions of \(\lambda_\rightarrow^?\). All type errors that cannot be statically resolved by the gradual type system are delegated to runtime checks.

Type inference allows programmers to omit type annotations in their programs and have the compiler infer the types for them. Hindley-Milner type inference is often cast as a two step process that consists of generating constraints and then solving them by a unification algorithm~\cite{wand1987inference,pottier2005moderneye,pottier2005essence}. The inference algorithm models the typing rules as equations, called constraints, between type variables, while the unification algorithm computes a substitution \(S\), which is a mapping from type variables to types, such that for each equation \( \tau_1 = \tau_2 \), we have \( S(\tau_1)  = S(\tau_2)\).

S\&V introduce the gradually typed lambda calculus with type variables ( \(\lambda_\rightarrow^{?\alpha}\)), which is \(\lambda_\rightarrow^?\) extended with type variables, \(\alpha\). They define a new relation, \emph{consistent-equal} ($\simeq$), which extends the \emph{consistent} relation from \(\lambda_\rightarrow^?\) to treatment \(\alpha\). Fig.~\ref{fig:lam-dyn-a-ts} compares the typing rules for \(\lambda_\rightarrow^\alpha\), the statically typed lambda calculus with type variables, to the new type system \(\lambda_\rightarrow^{?\alpha}\). S\&V also specify a unification algorithm for \(\lambda_\rightarrow^{?\alpha}\) which integrates the \emph{consistent-equal} into Huet's unification algorithm~\cite{huet1975unification,knight1989unification} which is a popular algorithm that doesn't rely on substitution.

Huet's unification algorithm uses a graph representation for types. For example, a type like $Int \rightarrow \beta$ is represented as a sub graph in Fig.~\ref{fig:huet}. A node represents a type, ground types, type variables or the function type ($\rightarrow$), and edges connect the nodes of types belonging to a $\rightarrow$ type. From this it follows that the unification algorithm is the amalgamation of two graphs present in a constraint equation following the rules of the type system. Huet's algorithm maintains a union find structure~\cite{tarjan1975efficiency} to maintain equivalence classes among nodes and thereby types. When node \(A\) unifies with node \(B\) according to the type rules, the merge results in one of the two nodes becoming the representative of the merge.  This signifies that the representative node is the solution to the constraint being unified.  Fig.~\ref{fig:huet} shows how the unification of the constraint \( \{ \alpha \rightarrow \alpha = Int \rightarrow \beta  \} \) proceeds.

\section{Introduction to BCL}
\label{sec:bcl}
Our motivation to explore gradual types for BCL is rooted in several historical and contextual details, which we discuss in this section.  It is first helpful to understand that BCL is predominantly used to model financial contracts, by providing end users with programmatic access to a financial contract library.  The library we use is based upon the composable contracts of Peyton Jones, Eber and Seward~\cite{jones2000contracts}. Its internal contract data structure is used throughout our broader derivatives system to support various downstream analyses.  In this way, BCL serves as an expressive front-end for describing contracts to our derivatives system.

\lstset{,morekeywords={with,of,if,then,else,let,in,match} }

\begin{figure}

    \centering
    \begin{lstlisting}[mathescape=true,basicstyle=\small\ttfamily]
    
    let receive currency amount = scale (one currency) amount in

    let european_stock_option args =
      let first = stock_price args.effective_date args.company in
      let last = stock_price args.expiry_date args.company in
      let payoff = match args.call_or_put with
        | Call -> (last / first - args.strike)
        | Put -> (args.strike - last / first)
      in
      european args.expiry_date (receive args.currency payoff)
    in
    
    european_stock_option
      { company = "ABC Co.",
        call_or_put = Call,
        strike = 100.0,
        currency = USD,
        effective_date = 2021-01-17,
        expiry_date = 2021-01-22 }
        
    \end{lstlisting}
    
    \caption{European stock option}
    \label{fig:euro_stock_opt}
    
\end{figure}

Let us look at a short illustrative example of BCL code. Fig.~\ref{fig:euro_stock_opt} provides an example of the sort of thing for which BCL might be used. The \TT{european\_stock\_option} function produces a \TT{Contract} which models a European stock option.  European stock options grant their holder the right, but not the obligation, to buy or sell stock in a company.  The ``European'' in European stock option refers to the fact that, on one specific date, the holder must choose whether or not s/he would like to buy (or sell) the stock.  This is in contrast to ``American'' options, where the holder may choose to buy (or sell) on any date within a specified range of dates.

This stock option is based on several helper functions, defined in~\cite{jones2000contracts}, which we must examine first. The \TT{european} function constructs a contract which allows its holder to choose between receiving ``something'' or nothing on a specified date. \TT{receive} constructs a contract that pays the specified \TT{amount} of the specified \TT{currency} passed as arguments andn uses the \TT{scale} and \TT{one} primitives. The \TT{scale} primitive takes an amount of type \(Obs\ Double\)--where type \(Obs\ d\) represents a time-varying quantity of type \(d\)--and a contract as arguments and multiplies key values in the contract by the amount.  Note that \TT{european\_stock\_option} uses \TT{-} and \TT{/} operators which are built-ins that operate on \(Obs\ Double\) arguments.  \TT{stock\_price} is a  primitive for looking up the price of the specified stock on the specified date.

\TT{european\_stock\_option} starts off by using \TT{stock\_price} to look up the price of the specified company's stock on the ``effective'' (contract start) and ``expiry'' (contract end) dates.  It uses these stock prices to construct the payoff based on the specified call or put style, and feeds the payoff to \TT{receive} to construct a contract that pays it.  Finally \TT{european\_stock\_option} passes the result of \TT{receive} to the \TT{european}, which allows the holder to choose between the payoff and nothing.  Note that the payoff may well be negative, so the holder's choice is not entirely clear.
The end of Fig.~\ref{fig:euro_stock_opt}, provides an example call \TT{european\_stock\_option} which constructs a call option on ABC Co.  In practice, functions like \TT{european\_stock\_option} would be defined in BCL's standard library, and would be called by users who wish to model European stock options directly or who wish to model contracts that contain such options as sub-contracts.

\subsection{Motivation for gradual types}
Given that BCL is mostly used to describe financial contracts, it should come as no surprise that our users are largely financial professionals.  In particular, many are financial engineers or quantitative analysts with some programming experience in dynamic languages such as Python and MATLAB.  Typically these users need to translate term sheets, plain-English descriptions of a contract, into BCL for consumption by our system.  These contracts are mostly one-off and, once finished, are unlikely to be reused as subcontracts to build further contracts.  For these reasons, the users of BCL are primarily concerned with development speed.  Ideally, they would like to be able to translate a term sheet as quickly as possible, so that they may focus on analyzing the contract's behavior once it has been ingested by our system.

On the other hand, the maintainers of BCL and its standard library are software engineers and functional programmers with extensive experience in OCaml, {\CC} and other static languages.  The main jobs of the BCL maintainers are implementing language extensions and standard library functions.  One of the significant constraints that they face is preserving backwards compatibility. All existing user contracts must continue to work as BCL evolves--even minor changes in behavior are unacceptable!  Given the broad reuse of the features that BCL's language maintainers implement and the difficulties involved in rolling back features, correctness is the paramount concern of BCL maintainers.

Finally, it is important to note that the version of BCL described here is actually the second version of BCL.  The first version of BCL was dynamically typed, so we will distinguish it from the second version by referring to it as Dynamic BCL.  Dynamic BCL supports only a few primitive data types, as well as a list composite type; it does not support algebraic types.  It also runs only minimal validation before attempting evaluation.  This simplicity makes Dynamic BCL well suited to our users who seek to quickly feed contracts into our system, but ill-suited to the library code written by our maintainers.  Additionally, some users who encounter runtime type errors while implementing particularly complex contracts would turn to the maintainers for assistance, further increasing the burden on the maintainers.  It was in light of these issues, that we developed (Static) BCL.

To address the issues with Dynamic BCL while remaining useful to our users, BCL aims to be a static language that feels roughly dynamic.  To this end, BCL supports implicit static types via type inference; we chose Hindley-Milner style inference so that our users could omit type annotations in almost all cases.  BCL also supports record and variant types, although they are structural rather than the nominal ones typically seen in OCaml and Haskell.  This choice also lends BCL a more dynamic feel.

The goal of BCL's design is to retain enough flexibility for our users, while introducing static types for the benefit of our language maintainers.  However, ``enough flexibility'' is entirely subjective and some users may well feel that any amount of static checking results in a system that is too inflexible.  Gradual types address this concern by allowing users to use dynamic types where they like, while also allowing maintainers to use static types where they would like.  Importantly, gradual types guarantee that fully dynamic code and fully static code can co-exist, and that static code is never blamed for runtime type errors.  Taken together, these two guarantees satisfy both groups, and ensure that the type errors that dynamic users see are isolated to the code that they themselves wrote.

\subsection{Core calculus}

BCL's core calculus is the lambda calculus extended with structural composite types and recursive types. Furthermore, BCL is implicitly-typed and supports Hindley-Milner style type inference.  This section describes the types and terms of this core calculus.  Note, however, that the grammars in this section are abstract representations of BCL's theoretical underpinnings, and do not cover the full set of productions in BCL's grammar.

\subsubsection{Kinds and Types}

\begin{figure}
\(\begin{array}{l c l l cl}
    \kappa &::=& * \mid \rho \mid \kappa \Rightarrow \kappa
&
    C &::=& \rightarrow \mid \Pi \mid \Sigma \mid ...
\\
    \tau &::=& \alpha \mid C \mid \tau\ \tau \mid l : \tau; \tau \mid \epsilon \mid \mu \alpha . \tau \quad
&
    \sigma &::=& \tau \mid \forall \alpha . \sigma
\end{array}\)
    \centering
    \caption{Grammar of types and kinds}
    \label{fig:typesandkinds}
\end{figure}

The grammar of the types and kinds that describe BCL is given in Fig.~\ref{fig:typesandkinds}.  Our kind system is fairly standard and consists of only three forms.  The base kind, \(*\), is the kind of ``proper'' types--\(Int\) and \(Int \rightarrow Int\), for example--which themselves describe terms.  The row kind, \(\rho\), is of course the kind for rows.  The operator kind, \(\Rightarrow\), is the kind of type operators --  \(Array\) and \(\rightarrow\), for example -- which take types as arguments and which do not directly describe terms.

\(C\) ranges over type constructors, including the type operators for function types (\(\rightarrow\) of kind \(* \Rightarrow * \Rightarrow *\)), record types (\(\Pi\) of kind \(\rho \Rightarrow *\)) and  variant types (\(\Sigma\) of kind \(\rho \Rightarrow *\)). \(C\) may also include additional constructors for base types (e.g. \(Int\) and \(String\)) and more type operators (e.g. \(Array\)) as desired.  However, these additional constructors are not useful for our purposes here, so we make no further mention of them.

Our type system is stratified into monomorphic types and type schemes, per \cite{damas1982principal}.  Monomorphic types, \(\tau\), consist of type variables, type constructors, and record, variant and recursive types.  Type variables are ranged over by \(\alpha\), \(\beta\), \(\gamma\), etc., and are explicitly bound by \(\mu\) and \(\forall\) types, as described below. Rows are written \(l : \tau; \tau'\), indicating that the row has a field labeled \(l\) of type \(\tau\).  \(\tau'\) has kind \(\rho\) and dictates the other fields that the row may contain.  If \(\tau'\) is a type variable, the row can contain arbitrary additional fields; if \(\tau'\) is the empty row, \(\epsilon\), the row contains no additional fields; finally if \(\tau'\) is another type of the form \(l : \tau; \tau'\), then the row contains exactly the fields specified therein. Recursive types are written \(\mu \alpha . \tau\), where the variable \(\alpha\) represents the point of recursion and is bound within \(\tau\).  BCL's recursive types are equi-recursive, so it does not have explicit constructs for rolling and unrolling recursive types.  Finally, type schemes have two forms: monomorphic types and universally quantified schemes.  Monomorphic types, \(\tau\), are merely the types described above.  Universally quantified schemes, \(\forall \alpha . \sigma\), bind the variable \(\alpha\) within the scheme \(\sigma\).  Naturally, it is through universal quantification that BCL supports parametric polymorphism.

\subsubsection{Terms}

\begin{figure}

\(\begin{array}{l c l}
  t &::=& \x \mid \lambda \x . t \mid t\ t \mid  {\tlet\ \rec\ \x = t\ \tin\ t}
  \mid  t : \tau \mid \{\overline{l_i : t_i}\}

  \mid   t . l \mid l\ t  \mid \match\ t\ \with\ \overline{l_i x_i \Rightarrow t_i}

\end{array}\)
    \centering
    \caption{Grammar of terms}
    \label{fig:terms}
\end{figure}

The grammar of the terms in BCL is given in Fig.~\ref{fig:terms}.  Most of the term forms are drawn directly from the lambda calculus.  Term variables are ranged over by \(x\), \(y\), \(z\), etc., and are introduced by lambda abstraction, let-bindings and match-expressions.  Lambda abstraction is written \(\lambda \x . t\) and binds \(x\) within the expression \(t\).  Lambda abstractions are eliminated by application, which is denoted by juxtaposition: \(t\ t\).  Let-bindings are written \(\tlet\ \rec\ \x = t\ \tin\ t\).  The \(\rec\) is optional and, when present, indicates \(x\) may be referenced by the expression to the right of the \(=\); \(x\) may of course always be referenced by the expression to the right of the \(\tin\).  Type annotations are written \(t : \tau\), and serve to ensure that \(t\) has the they \(\tau\).

In addition to the forms described above, BCL supports records and variants.  Record introduction is written  $\{\overline{l_i : t_i}\}$ , where \(t_i\) evaluates to the value stored in the field \(l_i\).  Records are eliminated by field projection.  The projection of the field \(l\) from the record \(t\) is written \(t . l\).  Variant introduction is written \(l\ t\), where the label \(l\) is used to tag the variant's payload, \(t\).  Variants are eliminated by case analysis, which is written \(\match\ t\ \with\ \overline{l_i x_i \Rightarrow t_i}\), which evaluates to the branch specified by the tag associated with the variant \(t\).

\section{Implementation}
\label{sec:implementation}
We identify three main components required to add gradual typing to a statically typed language with type inference, such as BCL. The first is the ability to annotate terms with types, as these annotations dictate whether a term is type-checked statically or dynamically. The second is the addition of a dynamic type to the existing set of types, and the third is an algorithm to unify the existing types with the newly added dynamic type. Since our grammar, shown in Fig.~\ref{fig:terms}, already supports explicit annotation of terms, we have the means to differentiate between dynamically typed and statically typed code. We add a dynamic type, \(\dyn\), to our set of types; it serves to indicate that a term that will be dynamically typed.  BCL's type inference algorithm statically infers a type for every term, meaning that by default BCL programs are completely statically typed. In order to tell the type system to dynamically type some terms, we must explicitly annotate those terms with the \(\dyn\) type.

For example: a simple increment function can be defined in BCL as follows.

\begin{lstlisting}[basicstyle=\small\ttfamily]
    let incr x = x + 1 in incr
\end{lstlisting}

The type system will infer the type \(Int \rightarrow Int\) for the \TT{incr} function.  However, we can instead provide an explicit annotation.

\begin{lstlisting}[mathescape=true,basicstyle=\small\ttfamily]
    let incr x = x + 1 in incr : $\dyn$ -> Int
\end{lstlisting}

In this case, the inference algorithm retains the annotated type as the type of the function.  Any type checks on the argument of the \TT{incr} function would be put off until runtime.
While the type checks pertaining to \(\dyn\) types are delayed, we still need to complete the inference procedure in order to infer the types of the un-annotated portions of the program (like the return type of \TT{incr}).  Siek and Vacchrajani~\cite{siek2008gradual}(S\&V) extend the standard unification-based inference algorithm to handle the \(\dyn\) type.  Their algorithm is based on the \emph{consistent-equal} relation which takes into consideration the type variables that are generated as part of a typical type inference algorithm. Fortunately for us, their algorithm works well for our implementation with only minor adaptations.

\lstset{emph={find,unify,infer,merge,copy\_dyn,maybe\_copy\_dyns,new,fresh\_type\_variable,new,was\_copied,was_visited,mark_visited,},emphstyle=\texttt,morekeywords={case,of,if,then,else,error,and} }

\begin{figure}

    \centering
    \begin{lstlisting}[mathescape=true,basicstyle=\small] 
    maybe_copy_dyns ($\tau_1 \simeq \tau_2$) =
      $\tau_1'$ $\leftarrow$ if was_copied $\tau_1$ then $\tau_1$ else copy_dyn $\tau_1$
      $\tau_2'$ $\leftarrow$ if was_copied $\tau_2$ then $\tau_2$ else copy_dyn $\tau_2$
      $\tau_1' \simeq \tau_2'$

    unify $\tau_1''$ $\tau_2''$ =
      $\tau_1$ $\leftarrow$ find $\tau_1''$
      $\tau_2$ $\leftarrow$ find $\tau_2''$
      if was_visited $\tau_1$ and was_visited $\tau_2$ then
        ()
      else case maybe_copy_dyns ($\tau_1 \simeq \tau_2$) of
        $\phantom{\mid}$ $\alpha \simeq \tau$ $\mid$ $\tau \simeq \alpha$ $\Rightarrow$ merge $\tau$ $\alpha$ (*Case 1 & 2*)
        $\mid$ $? \simeq \tau_1\ \rightarrow \tau_2$ $\mid$ $\tau_1\ \rightarrow \tau_2 \simeq\ ?$ $\Rightarrow$ (*Case 3 & 4*)
          unify $\tau_1$ (new $?$)
          unify $\tau_2$ (new $?$)
        $\mid$ $? \simeq \tau$ $\mid$ $\tau \simeq\ ?$ $\Rightarrow$ merge $\tau$ $?$ (*Case 5 & 6*)
        $\mid$ $\tau_{11}\ \rightarrow \tau_{12} \simeq \tau_{21}\ \rightarrow \tau_{22}$ $\Rightarrow$ (*Case 7*)
          unify $\tau_{11}$ $\tau_{21}$
          unify $\tau_{12}$ $\tau_{22}$
        $\mid$ $l : \tau_1; \tau_2 \simeq l' : \tau_1'; \tau_2'$ if $l$ = $l'$ $\Rightarrow$ (*Case 8*)
          unify $\tau_1$ $\tau_1'$
          unify $\tau_2$ $\tau_2'$
        $\mid$ $l : \tau_1; \tau_2 \simeq l' : \tau_1'; \tau_2'$ $\Rightarrow$ (*Case 9*)
          $\alpha$ $\leftarrow$ fresh_type_variable()
          unify ($l : \tau_1; \alpha$) $\tau_2'$
          unify ($l' : \tau_1'; \alpha$) $\tau_2$
        $\mid$ $\mu \alpha . \tau \simeq \tau'$ $\mid$ $\tau' \simeq \mu \alpha . \tau$ $\Rightarrow$ (*Case 10 & 11*)
          mark_visited ($\mu \alpha . \tau$)
          unify $\tau[\mu \alpha . \tau / \alpha]$ $\tau'$
        $\mid$ $\epsilon \simeq \epsilon$ $\Rightarrow$ () (*Case 12*)
        $\mid$ _ $\Rightarrow$ error

    infer $\Gamma$ $t$ =
      case $t$ of
        ...
        $t : \tau$ $\rightarrow$
          case (unify (infer $\Gamma$ $t$) $\tau$) of 
             Error $\Rightarrow$ Error : inconsistent types
            $\mid$ _ $\Rightarrow$ $\tau$
    \end{lstlisting}
    \caption{Type inference algorithm}
    \label{fig:type_inference_algo}
\end{figure}

Fig.~\ref{fig:type_inference_algo} shows an outline of our adaptation of S\&V's inference algorithm. Unlike the original algorithm by S\&V, BCL's does not separate constraint generation and constraint solving.\footnote{Put another way, our inference algorithm solves each constraint immediately after generating it, and before generating the next constraint.}  This difference is important, as it means that our inference algorithm does not have access to the whole constraint set prior to unification.  Instead the \TT{infer} function traverses the term, generating and solving constraints on the fly. For example, if it encounters an application \(t_1\ t_2\), it figures out the type of the term from the environment ($\Gamma$) and generates a constraint like \(\{ \tau_1 \simeq \tau_2 \rightarrow \alpha \}\), where \(\tau_1\) is the type of term \(t_1\), \(\tau_2\) is the type of \(t_2\) and \(\alpha\) is a fresh type variable.  \TT{infer} sends this constraint to \TT{unify}, which attempts to satisfy it or raises an error if the constraint cannot be satisfied.

Fig.~\ref{fig:type_inference_algo} shows the \TT{infer} case for a term \(t\) annotated with the type \(\tau\).  \TT{infer} generates a constraint which tries to unify the type inferred for \(t\) with the annotated type, \(\tau\). We highlight this case for two reasons.  First, the only way we can currently introduce a \(\dyn\) type in BCL is through an annotation.  Therefore, this is the only point where constraints involving the \(\dyn\) type originate.  Second it is critically important that this case returns the annotated type and not the inferred type. Note that in \TT{incr} the inferred type \(Int \rightarrow Int\) differs from--but is \emph{consistent-equal} with--the annotated type \(\dyn \rightarrow Int\).  We always want the user's explicit annotation to take precedence in this situation.

BCL's unification algorithm is already based on Huet's unification algorithm, which makes adopting the changes suggested by S\&V easier.  The crux of S\&V's algorithm lies in the way the \(\dyn\) type unifies with other types, and particularly with type variables.  When \(\dyn\) unifies with a type variable, S\&V's algorithm makes \(\dyn\) the representative node.  However, when \(\dyn\) unifies with types other than type variables, the other type becomes the representative element of the resulting set.  The \TT{find} and \TT{merge} functions in Fig.~\ref{fig:type_inference_algo} come from the union-find data structure that underlies Huet's unification algorithm.  Respectively, they compute a node's representative element, and union two nodes' sets keeping the representative element of the first node's set.

The first six cases of the \TT{unify} function handle unification with the \(\dyn\) type as laid out by S\&V. We say first six because Cases 1 and 2 take care of unifying the $\dyn$ type with type variables as specified by S\&V's algorithm. Cases 3 and 4 handle an edge case in their algorithm.  These two cases simulate the operational semantics of Siek and Taha~\cite{siek2006gradual}, which require constraints like \( \{ \dyn \simeq \alpha \rightarrow \beta \} \) to be treated as \( \{ \dyn \rightarrow \dyn \simeq \alpha \rightarrow \beta \} \). We use \TT{new} to create a new node different from what was passed in to handle this case. 

Cases 8-11 take care of unifying with row and recursive types, neither of which are covered by S\&V's solution.  However, it is our observation that these types do not require special handling. A constraint like \( \{  x : Int; \epsilon  \simeq \dyn \} \) would be handled by Case 2 and \(\dyn\) would be merged with the row type \(x : Int; \epsilon\). Now suppose the \(\dyn\) is present inside the row type like in the following constraint \( \{ x : \dyn; \epsilon  \simeq  x : Int; \epsilon  \} \); this will be handled by Case 8 and then Cases 5 and 12 when we recursively call unify with the types within the row.  
The same holds true for unification with the recursive type.  For example, a constraint like \( \{ List\ Int \simeq List\ \dyn \} \) will have the following successful unification trace:
{\footnotesize
\begin{align*}
    & \{ List\ Int \simeq List\ \dyn \}\tag{Case 10}  \\
    & \rightarrow \{ \Pi (head : Int; tail: List\ Int;\epsilon) \simeq List\ \dyn \}\tag{Case 11} \\
    & \rightarrow \{ \Pi (head : Int; tail: List\ Int;\epsilon) \simeq \Pi (head : \dyn; tail: List\ \dyn;\epsilon) \}\tag{Case 8} \\
    &\rightarrow \{ \{ Int \simeq \dyn \}, \{ List\ Int \simeq List\ \dyn \} \}\tag{Case 6, Case 10} \\
    & \rightarrow \cdots \\
    & \rightarrow \{ \Pi\ \epsilon \simeq \Pi\ \epsilon \}\tag{Case 8} \\
    & \rightarrow \{ \epsilon \simeq \epsilon \} \rightarrow ()\tag{Case 12}
\end{align*}
}%

Where \(List\) is defined as follows.

\( List\ \alpha \equiv \mu a.\Sigma (Nil : \Pi \epsilon; Cons : \Pi ( head : \alpha; tail : a; \epsilon); \epsilon)  \)
Note that BCL supports \textit{equi-recursive types}, as mentioned in Section~\ref{sec:bcl}, so \TT{unify} tracks the types it visits with \TT{mark\_visited} and \TT{was\_visited} to detect cycles.

\subsubsection{The \TT{copy\_dyn} conundrum:} The \TT{copy\_dyn} function is a crucial part of the way \(\dyn\) unifies with other types.   In S\&V's presentation, \texttt{copy\_dyn} ensures that each \(\dyn\) node in the constraint set is physically unique.  Without this step, multiple types might unify with the same \(\dyn\) node, and then transitively with each other.  This has the potential to cause spurious failures in \TT{unify}.  S\&V's solution to this is to traverse the constraint set and duplicate each \(\dyn\) node prior to unification; this is performed by their implementation of \TT{copy\_dyn}.  Unfortunately, we do not have access to the full constraint set, because our inference algorithm generates and solves constraints in one step.

Our first attempt at working around this issue was to call \texttt{copy\_dyn} as the first step in unification.  However, this leads to over copying. For example, consider the constraint \{ \(\dyn \rightarrow \alpha \simeq \alpha \rightarrow \tau \) \}. According to Case 7 of \TT{unify}, when \(\alpha\) unifies with the \(\dyn\) node, \texttt{copy\_dyn} is called and a new \(\dyn\) node is created in the union-find structure. But when \(\alpha\) then unifies with \(\tau\), \TT{find} looks up \(\dyn\) as \(\alpha\)'s representative element, and \texttt{copy\_dyn} is called once more. \(\tau\) therefore unifies with the new \(\dyn\) node, instead of the one which unified with \(\alpha\).  Thus, we lose the fact that \(\tau\) and \(\alpha\) are the same type.

To rectify this, we implement \TT{maybe\_copy\_dyns}, which traverses a constraint and copies each \(\dyn\) node exactly once.\footnote{There are many ways to accomplish this.  Our approach was to use one canonical \(\dyn\) node in type annotations, and compare each \(\dyn\)'s address to the canonical node's address before copying.}  The result of this is the same as originally intended by S\&V's \TT{copy\_dyn} function.  That is, we ensure there is a unique \(\dyn\) node in the union-find structure for every unique use of \(\dyn\). 

\subsection{Discussion}
In Section~\ref{sec:background} we gave an overview of how statically typed languages approach this problem of promoting dynamic typing.
It is our observation that most statically typed, object-oriented languages approach dynamic typing following Abadi et al. That is, their dynamic type exploits subtype polymorphism to bypass static type checking.  This is a natural direction for object-oriented languages which rely heavily on subtyping. In order to inspect the types at runtime, these languages make use of type reflection.  Java is one such language where work has been done to add dynamic types using reflection, contracts and mirrors~\cite{gray2005fine}.  The Java Virtual Machine supports many dynamic languages like Jython and JRuby, demonstrating that such runtime constructs help static languages add more dynamic type checks. However, these implementations only add dynamic checks, and do not achieve the gradual goal of a seamless mix of static and dynamic types as in ~\cite{siek2015criteria}.  To our knowledge, only Featherweight Java~\cite{10.1145/503502.503505} has attempted to support proper gradual typing~\cite{ina2009gradual}. In any case, the primary purpose for dynamic types in these languages is inter-operation with other dynamic languages.  This differs from our own purpose and the end result does not fit our needs well.  Thus we conclude that this approach was not a good design choice for us. 

The languages closest to BCL are statically typed functional languages with type inference, such as SML, OCaml, and Haskell. OCaml has incorporated dynamic typing at the library level by leveraging its support for generalized algebraic data types~\cite{balestrieri2018generic}.  Similarly, Haskell supports a dynamic type as a derivative of the \TT{Typeable} type class, which uses reflection~\cite{jones2016reflection} to look at the runtime representation of types. While these approaches introduce more dynamism, they lack the simplicity of gradual typing, which hide all the nuts and bolts of the type system under a simple $\dyn$ annotation.

Seamless interoperation of static and dynamic types as promised by gradual typing fits very well with our use case. It lets our end users access both paradigms without knowledge of specialized types or constructs.  Furthermore, the approach we use---extending unification-based inference with gradual typing---is a natural extension for languages like BCL, which support static type inference. The addition of dynamic types to the type system easily boils down to how we handle this new type in the unification algorithm, and does not require reworking the entire type system. We attribute this benefit to S\&V's proposed inference algorithm, which incorporates the essence of the $\lambda_\rightarrow^{?}$ type system.  This makes it easier to adapt to an existing language with similar constructs.

Garcia and Cimini's work takes a different approach to this problem but their end goal is the same: gradual type inference in a statically typed language. The authors of that work feel that S\&V's approach has ``complexities that make it unclear how to adopt, adapt, and extend this approach with modern features of implicitly typed languages like let-polymorphism, row-polymorphism and first class polymorphism''. Our experience with S\&V's approach was different: we found the integration fairly simple without major changes to the original inference algorithm. We leave a deep dive into the differences between these two schemes to future work. Based on Garcia and Cimini's design principle,  Xie et al.~\cite{10.1145/3310339} introduce an inference algorithm with support for  higher-rank polymorphism, using a \emph{consistent subtyping} relation. In contrast, BCL only infers rank-1 polymorphic types and doesn't support higher-rank polymorphism.

We recognize an added benefit of going from a static to a dynamic language with explicit \(\dyn\) annotations. Promoting static type checking in a dynamic language without type inference requires the programmer to add annotations to all parts of the code that they want statically checked. Needing to add these annotations is such a burden for the programmer that they often skip some annotations and miss out on static optimizations. These un-annotated types are implicitly dynamic, leading to runtime overhead, despite the fact that on many occasions they could be statically inferred.  This in turn has lead to efforts to making gradual typing more efficient~\cite{10.1145/2103621.2103714}.

BCL does not have this issue as it provides static inference by default.  It therefore enjoys the optimizations of static typing and and can skip unnecessary runtime checks. Moreover, BCL could support a dynamic-by-default mode with an additional compiler flag that implicitly annotates un-annotated terms with the $\dyn$ type.  This  makes it even more seamless to go from complete static typing to complete dynamic typing. We might also consider doing this implicit annotation on a file-level or function-level.  In cases where it is possible to separate dynamic and static components, this could even lead to cleaner refactoring. These ideas have not yet been implemented in BCL but are something we intend to do as future work.

\section{Application of gradual types}
\label{sec:benefits}
Gradual typing enables the quick prototyping common in dynamic languages, as well as specialized applications that enable simplification of  existing code.  In this section, we focus on the latter, due to space constraints.  Notice, in Fig.~\ref{fig:euro_stock_opt}, that the \TT{scale} combinator~\cite{jones2000contracts} is simply multiplication of a contract by a floating-point \emph{observable}. In a domain specific language like BCL, it is convenient to reuse the \TT{**} multiplication syntax for \TT{scale} as well.  We can fit this more general \emph{observable} multiplication operator into the type system with the gradual type \(Obs\ Double \rightarrow \dyn \rightarrow \dyn\).  Our new multiplication operator can delegate to \TT{scale} when the second argument a $Contract$ at runtime and continue with \emph{observable} multiplication, or raise a runtime type error based on the runtime type of the second argument.  With this new operator, the \TT{receive} function can be rewritten thus:
\lstset{morekeywords={with,of,if,then,else,let,in,match} }

\begin{lstlisting}[mathescape=true,basicstyle=\small\ttfamily]
let receive currency amount = one currency ** amount in ...
\end{lstlisting}

There are a variety of extensions to Hindley-Milner that enable this sort of ad-hoc polymorphism statically. Type classes, for example, extend the signature of overloaded functions with classes~\cite{wadler1989adhoc}, which our users would need to learn.  Similarly, modular implicits introduce a separate syntax for implicit modular arguments~\cite{white2015implicits}.  However, these constructs require effort to educate our end users in their use and detract from the dynamic feel of the language.  Gradual types, by contrast, are much easier for our end users since they already work in a dynamic environment and it does not require new syntax (save a $\dyn$ annotation).

It is worth noting that, while the new multiplication operator can be given a valid type in BCL, it cannot currently be implemented in BCL; it can only be implemented as a built-in operator because BCL provides no way to perform case analysis on the runtime type of a dynamic value.  However, addressing this is actually quite easy if we reuse BCL's existing support for variants.  That is, we could implement a \TT{dynamic\_to\_type} primitive which consumes a dynamic value and produces a variant describing its runtime type.  This would allow us to then branch on this variant with the existing \(\match\) construct. Fig.~\ref{fig:dynamic-mul} shows a prototype of a function that achieves this effect assuming the \TT{dynamic\_to\_type} primitive is defined. 

\begin{figure}
    \centering
    \begin{lstlisting}[mathescape=true,basicstyle=\small\ttfamily]
    let dyn_obs_mul x y = match dynamic_to_type (y) with 
      | Obs Double => x ** y
      | Contract => scale x y 
    in
        dyn_obs_mul : Obs Double $\rightarrow$ Dyn $\rightarrow$ Dyn
    \end{lstlisting}
    \caption{Sample of a dynamic Observable multiplication function}
    \label{fig:dynamic-mul}
\end{figure}
\TT{dynamic\_to\_type} is interesting in light of our discussion in Section~\ref{sec:background}, which describes dynamic programming as the territory of BCL's users and not its maintainers.  Clearly, however, the dynamic multiplication operator is something that would live in BCL's standard library and be maintained by the language maintainers.  Indeed there are a number of interesting standard library functions which we might add on top of \TT{dynamic\_to\_type}.  Another simple example would be a \TT{any\_to\_string} function, which could produce a string representation for arbitrary types by traversing their runtime type and delegating to the appropriate type-specific to-string function.  Such a function would be very handy for debugging and quick inspection of values.

The \TT{any\_to\_string} example is a function which consumes an arbitrary value.  However, there are equally compelling use cases for producing arbitrary values.  For example, property-based testing frameworks rely on automatically generating values that conform to certain constraints.  We could implement a simple property-based testing framework with a function which consumes the output of \TT{dynamic\_to\_type} and generates arbitrary values that conform to that type.  Such a framework would be especially useful in a domain such as ours, where real money is at stake, and where robust testing is absolutely critical.

\section{Conclusion}
\label{sec:conclusion}
Dynamic languages are extremely popular with many users.  For users with a limited computer science background, for whom ease-of-use is the paramount, this is doubly true.  However, despite the flexibility offered by dynamic typing, the safety offered by static typing is helpful in domains where correctness is critical.  In such an arena, gradual types are a perfect blend of both paradigms, and they provides a middle ground to please a larger group of users.  Given this, it is important for the literature to speak about adapting gradual types to existing languages.
As a first step towards that, we write about our experiences adapting gradual typing to our implementation of a statically typed functional language with type inference.  We provide context in terms of how others in similar situations approached this problem, and we elaborate our inference algorithm with key insights around what worked for us and what did not. 
We identify an interesting use case for gradual types here at Bloomberg, where we look to harmonize end users and language maintainers with competing goals. End users want to specify financial contracts without worrying about static typing demands, while language maintainers need a more rigorous type system that ensures that libraries that they write are error-free. Gradual types allow us to satisfy both groups. 
We also intend to gather feedback from our end users and maintainers about how gradual types are being used , which can give insight into possible tweaks to make this system more amenable to all.

%
%
%

\bibliographystyle{splncs04}
 
\bibliography{main}

\end{document}